# Probing the Temperature Profile of Energy Production in the Sun

## Christian Grieb and R. S. Raghavan

*Institute of Particle, Nuclear and Astronomical Sciences and Department of Physics*
*Virginia Polytechnic Institute and State University, Blacksburg VA 24061*

The particle kinetic energies of pp fusion in the sun (Gamow Energy) produce small changes in the *energies* of pp solar neutrinos relative to those due only to exothermal energetics. Observation of this effect may be possible via the unique tools of the upcoming LENS solar neutrino detector. The temperature profile of energy production in the sun may thus be directly probed for the first time.

Recent solar neutrino (ν) experiments have created a fundamentally new ν phenomenology and confirmed the prediction of the tiny ($< 10^{-4}$) $^8$B solar ν-flux. The new paradigm for solar ν research is: precision tests of critical details of particle and astrophysical theory inferred so far and discovery of new surprises. The focus is on low energies (<2 MeV) with >99% of the ν flux, but as yet, with no real-time spectroscopic data essential for the new paradigm. It is thus the new frontier for discovery.

In that spirit, we consider a new goal: a direct probe of the temperature (T) profile of the energy producing core of the sun. At the high T of the core ($\sim 15 \times 10^6$ K), thermonuclear fusion of 4p $\rightarrow$ He creates the energy released by the sun. Neutrinos emitted in the primary pp fusion and the subsequent reaction chain --the decays of $^7$Be and $^8$B--render this process uniquely observable to terrestrial ν detectors. Experiments so far have focused on ν *fluxes* to test predictions of the standard solar model (SSM)[1].The new goal focuses on precise *energies* and spectral shapes of ν's from the basic pp fusion.

Thermal kinetic energies needed for initiating pp fusion (Gamow energy) affect the energies of the emitted ν's relative to those expected from laboratory exothermal energetics only. Inclusion of the thermal energies changes the spectral shape of the pp ν continuum and its maximum energy. The kinetic energies of free *electrons* in the ionized solar plasma cause additional shifts of the mean energy of the electron capturing (EC) pep and $^7$Be ν lines[2]. The ν's originate in different inner shells in the sun at different temperatures.[3] Precise measurement of the *small* energy shifts, viz. [(ν energy in sun)–("lab" ν energy)] of pp, pep and $^7$Be ν's, thus probe the temperature and energy production profiles in the solar core directly. We note that pp fusion has never been observed in the laboratory. This paper explores the detectability of the Gamow energy shift in the spectral signal of pp ν's.

This goal is timely. Until recently, just the observation of pp solar ν's in real-time was thought to be an insurmountable technological challenge. Great progress has since been made to meet this challenge and measure low energy ν fluxes.[4] Even so, the measurement of solar ν *energies* with sufficient precision to reveal the small shifts above, demands more advanced technical tools. Those will be available in the upcoming LENS (Low Energy Neutrino Spectroscopy) detector[5]. The *tagged* ν-capture reaction of LENS yields a signal electron of specific energy $E_e = E_v – Q_d$ ($Q_d$ = threshold energy) that uniquely identifies the initial ν energy. Thus the incident ν spectrum (offset by $Q_d$) is directly obtained. The value of $Q_d$ is low enough that ~95% of the pp ν spectrum is accessible, a major need for the present goals. The reaction tag specifies not only the signal but independently allows a precise, *on-line* measurement of background. This combination of technology, achieved so far only in LENS, is indispensable for precision measurement of ν energies. We show that the LENS framework can lead to a definitive measurement of the pp Gamow energy and the temperature profile of energy production in a main sequence star for the first time.

We focus on three solar reactions in the sun, two from the primary pp fusion (ppI) and the third in the subsequent chain (ppII)[6]:

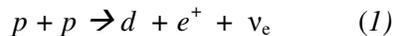
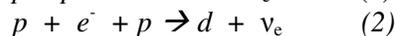
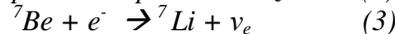

$$p + p \rightarrow d + e^+ + \nu_e \qquad (1)$$
$$p + e^- + p \rightarrow d + \nu_e \qquad (2)$$
$$^7Be + e^- \rightarrow\, ^7Li + \nu_e \qquad (3)$$

These are the only reactions with measurable effect of the ambient temperature on the ν energies. The 3-body reaction (1) produces a continuous $\nu_e$ spectrum. In (2), pp fusion occurs with EC from the solar plasma, a 2-body process that emits a mono-energetic ν line. In (3) $^7$Be decays by EC. Thus it also emits a ν–line. The ν energies of (1-3) based on "lab" energetics (difference in masses of initial and final states) are given in Table 1. The "lab" pp ν spectrum on this basis is given by:

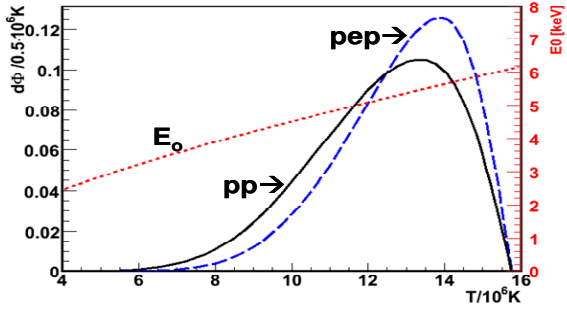

*Fig. 1 Fractional flux $d\varphi_{pp}(T)/dT$ (full line, black) and $d\varphi_{pep}(T)/dT$ (dashes blue) vs T; Gamow energy $E_o(T)$ (dots red) (right hand scale) vs T. The pp v production region with the above temperatures extends up to fractional solar radius R =0.5.*

$$P_{lab}(q,Q_s) \sim q^2 \, p \, W \, F(Z,W,Q_s) \qquad (4)$$

where p, W are momentum and energy of the e$^+$ in (1), F is the Fermi function with Z= -1 and $Q_s$ is the exothermic energy released = $(q_{max}-mc^2)$. The spectrum for reaction (1) is a continuum with q and W defined by q = $Q_s$–W; those for (2) and (3) are lines with a single q = $q_{max} = Q_s$.

$P_{lab}(q)$ does not take into account the particle kinetic energies necessary to achieve fusion. The average thermal energy for pp fusion (Gamow peak) $E_0(T) = 1.22[A (=0.5)]^{1/3} T^{2/3} = 5.9 \, T_{15}^{2/3}$ keV where $T_{15}$ is the ambient temperature in units of 15x $10^6$K.[7] T varies over the production region of pp v's. Fig. 1 shows $d\varphi_{pp}(T)/dT$ and $E_0(T)$ vs. T in the solar core derived from SSM data[8]. The effect of high *density* in the sun (~150g/cm$^3$) via the pycnonuclear (in addition to the *thermo*nuclear) effect[9] that could slightly change the three profiles in Fig. 1, is ignored. The pp v spectrum in the sun $P_{sun}$ is calculated as $P_{lab}(q,Q_s)$ with the Gamow energy $E_0(T)$ added to $Q_s$.[10] It is weighted by the fraction $d\varphi_{pp}(T)$ of the pp v flux produced at that T and summed over the range of T in the pp fusion region.

$$P_{sun}(q) = \sum_{\varphi_{pp}} P_{lab}(q, Q_s + E_0(T)) d\varphi_{pp}(T) \qquad (5)$$

The shape of $P_{sun}$ (q) is shown in Fig. 2 (upper panel). The *mean* energy in the accessible pp v spectrum (E>114 keV) shifts by $\Delta$<E> = 2.98 keV. The new *endpoint* of the pp v spectrum is $q_{max}$(sun) $\approx (q_{max}+\Delta E)$ with $\Delta$E= 5.2 keV (in agreement with eq. 38 of Ref. 10). The experimental detection of $\Delta$E (as appearing in Fig. 2) using fits to the signal from the spectrum $P_{sun}$(q) (5) probes the temperature profile of the pp fusion region. Spectral changes via several other effects were considered[10] and shown to be small. The pp v's are flavor-converted by vacuum oscillations that are energy independent. Matter effects with current v parameters create at worst a <0.3% effect. A pp v signal consistent with $P_{sun}$ (rather than $P_{lab}$) can thus be safely ascribed to the pp Gamow effect.

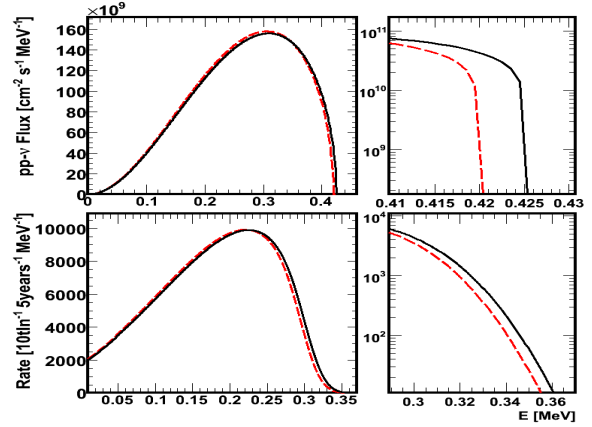

*Fig. 2 Spectral shape $P_{lab}(q)$ (dashed, red) and $P_{sun}(q)$(solid, black) (upper left). Detail of spectral shift (upper right). Electron signal spectra in LENS (background subtracted) with energy resolution of 6% (σ) at 300keV (lower left) (Ref. 5). Detail of spectral shift of signal (lower right);*

In the case of the $^7$Be v line, the decay occurs long after $^3$He+$^4$He fusion, thus the Gamow energy does not enter into the v energy. However, kinetic energies of the electrons in the $^7$Be EC decay do. The mean energy of the 861 keV Be line is shifted by $\Delta$<E>(Be) = 1.41$T_{15}$ keV. Averaged over the temperatures in the region of origin of Be v's, $\Delta$<E>(Be) = 1.29 keV[2]. Since kinetic energies are positive they always *add*, making the line asymmetric[2]. The pep v line results from pp fusion *and* electron capture. Thus the Gamow energy shift of the v energy (using $d\varphi_{pep}(T)$ from Fig. 1), is <$E_o$(T)$d\varphi_{pep}$(T)> = 5.35 keV. Further, the line becomes asymmetric as in the Be case. The total shift $\Delta$E (pep) =5.35+1.3 = 6.65 keV. The ("lab") exothermal energies and the expected thermal shifts of reactions (1-3) in the sun are given in Table 1.

The LENS experiment is based on the charged current (CC) driven $v_e$ capture in $^{115}$In[11]:

$$v_e + {}^{115}In \rightarrow e^- + {}^{115}Sn^* \rightarrow 2\gamma + {}^{115}Sn \qquad (6)$$

The $v_e$ capture leads to an isomeric state in $^{115}$Sn at 614 keV which emits a delayed (τ = 4.76 µs) cascade 2γ (= 116+498 keV) that de-excites to the ground state of $^{115}$Sn. The reaction threshold is $Q_d$ = 114(4) keV. The $v_e$ signal is the electron in (6) whose energy $E_e = Ev_e + Q_d$ leads to the incident $v_e$ energy. The delayed 2γ signal is a powerful tag for the $\nu_e$ capture. The only valid events in LENS are *coincidence* events, the time distribution of which is shown in Fig. 3 (top panel). An exponential decay fit (with the signature lifetime of $^{115}$Sn*) to the time spectrum leads to the true signal events that occur at early delays ~10µs and the background from random coincidences at long delays. The random coincidence background arises dominantly from the natural β decay of the target $^{115}$In with end point of 498(4) keV[12] as shown in Fig 3. The background at long delays is measured separately, concurrently and precisely (with a wide gate of ~100 µs on the delay time). Background *subtraction* errors are thus minimal.

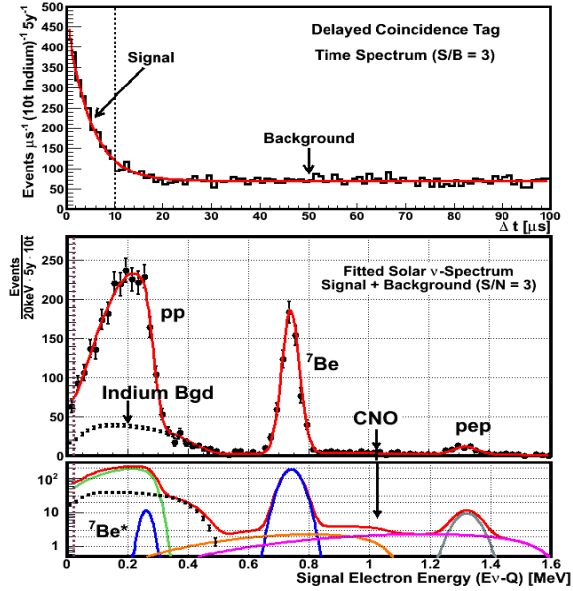

*Fig. 3 Signal spectra in LENS [5]. Top panel: delayed coincidence time spectrum fitted to isomeric lifetime τ=4.76 µs. Middle panel: energy spectrum at delays <10µs. Inset, random coincidences (from the pure β-decay of In target) at long delays. Bottom panel (in logarithmic scale): detail of weak CNO signal.*

The e-signal spectrum is:

$$N(E)dE \sim P_{sun}(q)WpF(Z,W)dq \qquad (7)$$

$W/mc^2 = E_e = (q - Q_d + 1)$, $p = \sqrt{(W^2 - 1)}$ and F is the Fermi function for Z = +50 (Sn). The signal energy spectrum (at short delays) is shown in the middle panel of Fig. 3. It is the complete low energy solar $\nu_e$ spectrum with well-resolved features from pp fusion (pp and pep) and the $^7$Be and CNO decays (see bottom panel). The relative intensities in Fig. 3 follow from the current (LMA) $\nu_e$ conversion model, and the analysis efficiencies in LENS[5]. The net pp signal spectra NE(dE) expected in LENS from the pp ν spectra $P_{lab}(q)$ and $P_{sun}(q)$ (in the upper panels of Fig. 2) are shown in the bottom panel of Fig. 2. These spectra (from (7)) include the smearing due to detector resolution. The right panels in Fig. 2 show detail of the observable Gamow spectral shift.

The pp ν energy shifts in Fig. 2 are *small* as expected. We have performed extensive Monte-Carlo simulations to assess the detectability of these shifts. Background subtracted spectra such as in Fig. 2 were simulated for the incident ν spectra $P_{sun}(q)$ (eq. 5) with ∆E=5.2 keV for various detector parameters, chiefly the mass of In target M, the S/B ratio, the exposure time t and the energy resolution ε = (photoelectrons pe/MeV). This spectrum was then fitted to (7) with ∆E and the incident ν flux as free parameters. The procedure was repeated $10^4$ times to obtain a distribution of the fit ∆E values. The standard deviation of this distribution is δE, the precision attainable for a ∆E that fits the entire shape of the experimental spectrum N(E)dE. Fig. 4 shows the ∆E distribution and a Gaussian fit to this histogram that yields δE obtained for the detector parameters: M=12 ton, S/B=5, t = 6y and ε = $10^3$ pe/MeV. We also estimated the precision δ<E> attainable for the *mean* energy in the pp spectrum and for the $^7$Be and pep lines. They are compared to theoretical expectations from $P_{sun}$ in Table 1.

The LENS detector is a novel "scintillation lattice chamber", an optically segmented, three-dimensional array of ~0.5ℓ cells of liquid scintillator loaded with ~8-10 wt% Indium[5]. The scintillation signal from each cell will be always viewed by the same set of 3-6 phototubes. Thus, the full scale LENS of ~125 tons, though large in size, is in essence, a large array of small detectors capable of bench-top precision nuclear spectroscopy.

The measured δ<E> and δE are limited by the signal statistics, the energy resolution ε and the systematic errors from the energy calibration and the value of the ν-capture threshold $Q_d$. The latter are key for measuring *absolute* energies.

The event response of LENS will be continually monitored using internal radioactive sources emitting energy standards of *electrons* similar to the solar signal electrons. A prime example is the high rate (<<In decay rate) internal organic $^{14}$C with a β-continuum up to 156.475 keV of precisely known spectral shape with closely similar response as the pp continuum up to 306 keV. Electron line standards such as $^{207}$Bi (30y) with $E_e$ = 481.61(6), 554.37(10), 975.57(6) and 1048.1(1) keV can be added to the scintillator at low levels compared to the $^{115}$In activity. Simulations show that with the high statistics of single events attainable from these sources, the lines and absolute energies can be fixed to a precision δE(cal) < 0.3 keV.

*Table 1. Neutrino energies and thermal shifts*

| | q (lab) keV | +Δ⟨E⟩ keV | +δ⟨E⟩ keV | +ΔE keV | + δE keV |
|---|---|---|---|---|---|
| pp | 420.2[a] | 3.41[b] | 1.6 | 5.2[c] | 1.7 |
| pep | 1442.2 | 6.65[b] | 4.54 | | |
| ⁷Be | 861.8 | 1.29[b] | 0.81 | | |

[a]Maximum energy;  [b]Shift of mean energy of signal spectrum in the detector, in the case of pp in the energy range ⟨110-340⟩ keV;  [c]Shift of maximum energy in sun. The δE includes likely systematic errors (see text)

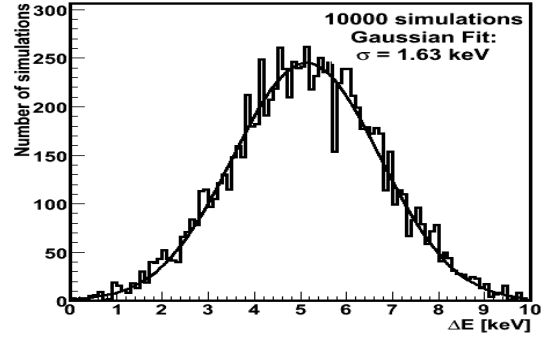

Fig. 4 *Distributon of the shift ΔE in the pp-ν maximum energy in 10000 simulations of randomly created signal spectra in LENS (Fig. 2 bottom panel). A Gaussian fit shown above yields δE = 1.63 keV (1σ)*

The $Q_d$ value can be fixed using 2 internal standards and an external one. The first is the sun itself, via the internal ⁷Be line with a measurable precision of δE = 0.81 keV (Table 1). The LENS project plans a ν -calibration (LENS-CAL) using a MCi source of well known monoenergetic ν's from ⁵¹Cr or ³⁷Ar whose signals in an Indium detector serve in same way as the solar ⁷Be line. The end-point of the ¹¹⁵In β-decay measurable in LENS with high accuracy, directly leads to $Q_d$. We expect δE($Q_d$) <0.3 keV using all three independent sources. The precision δE (pp) in Table 1 includes the systematic errors from δE(cal) and δE($Q_d$) added in quadrature.

The main result of Table 1 is the precision achievable in LENS for the energy shift δE(pp) in the pp ν maximum energy (last two columns). The predicted Gamow shift ΔE (Fig. 1 & 2) can thus be measured with ~95% confidence.

Astrophysically, the experimental value of ΔE can meaningfully probe the temperature profile of pp fusion shown in Fig. 1. For example, the shape of the profile can be changed in principle, without changing the integral pp ν flux. More generally, the full structure of the SSM predicts the following basically correlated observables: 1) the pp Gamow shift and the T-profile of pp fusion; 2) the pp ν flux $\propto$ (1-0.08(T/$T_{SSM}$)⁻¹·¹)¹³; 3) the ⁸B ν flux¹⁴ (via neutral & charged current detection) $\propto$ (T²⁵)¹³; and 4) the ν–luminosity¹⁵ $L_\nu$. The ultimate calibration of the SSM is: 5) the *photon* luminosity $L_{\odot}$, the true measure of the radiative energy released by the sun. With $L_\nu$ inferred from the global solar ν flux data, the ratio $L_\nu/L_{\odot}$ should be = 1.0 if the following conditions are met:  1) the ν model is correct; 2) the sun is in thermal quasi-equilibrium i.e. the energy generated in the interior *at present* ($L_\nu$) equals the radiative energy released ($L_{\odot}$) at present, some ~10⁵ *years after* generation; and 3) the sole source of the sun's energy is nuclear fusion. The best estimate from all solar ν data so far¹⁶, $L_\nu/L = (1.4^{+0.2}_{-0.3})$, offers little useful constraint. LENS can measure the Gamow shift as well as the pp ν flux and $L_\nu$ precisely (3-4%) with data of the quality of Fig. 3.

The precisely known $L_{\odot}$ completes the set of 5 observables. A global analysis of these observables allows a comprehensive new test of the SSM and the ν model. Initial results on ΔE and the interplay with this global analysis may motivate higher precision in the future. Our analysis shows that a x2 In mass could yield the precisions δE(pp) ~1.2 keV for the Gamow shift and δ⟨E⟩ (pp) ~1.1 keV, δ⟨E⟩(Be) ~0.57 keV and δ⟨E⟩(pep) ~3.1 keV for the mean energy shifts, that can usefully constrain the global analysis further.

We thank Mark Pitt and Hamish Robertson for helpful comments on the manuscript.